\documentclass[twocolumn,prl,aps,floatfix,superscriptaddress]{revtex4-1}

\usepackage{color}
\usepackage{ulem}
\usepackage{siunitx}

\usepackage{graphicx}

\usepackage{amsthm}
\usepackage{amsmath}

\newcommand{\ket}[1]{\vert #1 \rangle}
\newcommand{\eket}[1]{\bigl \vert #1 \bigr \rangle}
\newcommand{\bra}[1]{\langle #1 \vert}
\newcommand{\ebra}[1]{\bigl \langle #1 \bigr \vert}

\newcommand{\figref}[1]{Fig. \ref{#1}}
\renewcommand{\vec}[1]{\boldsymbol{#1}}
\newcommand{\R}{\boldsymbol{R}}

\begin{document}

\title{Particle entanglement in continuum many-body systems via quantum Monte Carlo}

\author{C. M. Herdman}
\email{Christopher.Herdman@uvm.edu}
\affiliation{Department of Physics, University of Vermont, Burlington, VT 05405, USA}

\author{P.-N. Roy}
\affiliation{Department of Chemistry, University of Waterloo, Ontario, N2L 3G1, Canada}

\author{R. G. Melko}
\affiliation{Department of Physics and Astronomy, University of Waterloo, Ontario, N2L 3G1, Canada}
\affiliation{Perimeter Institute for Theoretical Physics, Waterloo, Ontario N2L 2Y5, Canada}

\author{A. Del Maestro}
\affiliation{Department of Physics, University of Vermont, Burlington, VT 05405, USA}

\begin{abstract}
Entanglement of spatial bipartitions, used to explore 
lattice models in condensed matter physics, may be insufficient to 
fully describe itinerant quantum many-body systems in the continuum.
We introduce a procedure to measure the R\'enyi entanglement entropies on a {\it particle} bipartition, 
with general applicability to continuum Hamiltonians via path integral Monte Carlo methods.
Via direct simulations of interacting bosons in one spatial dimension, 
we confirm a logarithmic scaling of the single-particle entanglement entropy with the number of particles in the system.
The coefficient of this logarithmic scaling increases with interaction strength, saturating to unity in the strongly interacting limit.
Additionally, we show that the single-particle entanglement entropy is bounded by
the condensate fraction, suggesting a practical route towards its measurement in future experiments.
\end{abstract}
\maketitle

\begin{figure}
\begin{center}
\includegraphics[width=0.8\columnwidth]{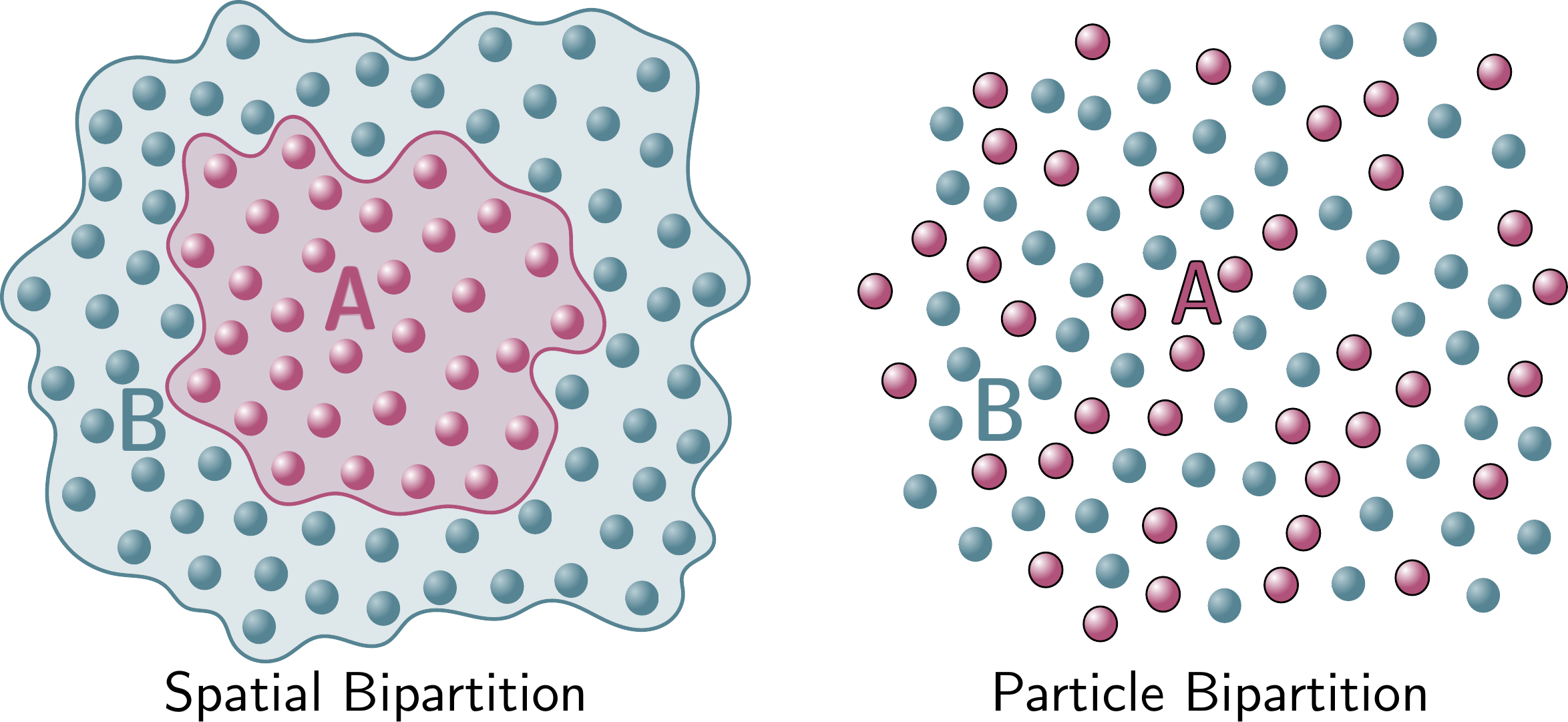}
\end{center}
\caption{(Color online). A comparison of spatial and particle bipartitions in the continuum.  Particle bipartitions are possible even in the case of their indistinguishability through a fictitious labeling scheme.}
   \label{fig:bipart}
 \end{figure}

Traditional two-point correlation functions, and their ability to probe broken symmetries, 
underlie our modern edifice of condensed matter theory.  However, they are known to fail as a foundation for a complete classification of all phases of quantum matter, as demonstrated spectacularly in fractional quantum Hall and other topological phases~\cite{Wen1990,Nayak2008}.    
To remedy this insufficiency, one can construct 
classifications based on information theory, which is by definition a complete description of all correlations, raising the question of {\it which} information-based quantities are relevant for quantum phases of matter~\cite{Horodecki2009}.
Bipartite entanglement entropy is a leading candidate, with its usefulness and versatility rapidly increasing along with an understanding of its properties in a variety of quantum phases~\cite{Amico2008}. 
For example, the ubiquitous ``area law'' in the spatial entanglement entropy of the ground state~\cite{Shredder,Eisert2010} 
has led to ways to classify and characterize quantum phases~\cite{Melko2010,Ju2012,Humeniuk2012} and 
phase transitions~\cite{Inglis2013b,Singh2011,Grover2012}  
in condensed matter systems,
along with elucidating the simulability of quantum models on classical computers~\cite{Latorre2007,Schuch2008a}.  
Previous work has primarily been
based upon {\it modal} bipartitions, where 
the entanglement is between two spatial or momentum subregions.
However, 
in systems of itinerant particles~\footnote{See Refs. \onlinecite{Balachandran2013,Benatti2012a,Wiseman2003} for examples of other measures of entanglement of identical particles.}, one can choose to bipartition into subsets of particles (\figref{fig:bipart})~\cite{Zanardi2002,Shi2003,Fang2003,Zozulya2008,Haque2009}.  This {\it particle} entanglement can give insight into not only quantum correlations due to interaction, but also exchange statistics and indistinguishability
\cite{Eckert2002,Haque:2007il,Zozulya2007a}.  

The utility of particle entanglement as a probe of quantum matter is illuminated by considering a spatially contiguous system of bosons. 
The non-interacting ground state is a Bose-Einstein condensate (BEC), which is a trivial product state in first-quantized notation. However, the entanglement entropy of a spatial subregion is nonzero, even in the absence of interactions, 
generated purely by number fluctuations~\cite{Simon2002,Vedral2003,Ding2009,Heaney2009,Song2010}. Conversely, in such a system, 
the particle entanglement is generated {\it solely} from interactions, and vanishes for the non-interacting BEC~\footnote{The particle entanglement can be non-zero in the absence of interactions for systems of fermions, as well as {\it disconnected} systems of bosons, where the entanglement arises from the exchange statistics alone. We focus on {\it contiguous} systems of bosons, where the ground state is a BEC}. In a lattice system of itinerant bosons, in the presence of strong interactions, the ground state will realize a Mott-insulator-like state which is an unentangled product state in a spatial basis, but has logarithmically extensive particle entanglement. Clearly, the interplay between spatial and particle entanglement has the potential to give insight into the nature of quantum phases of itinerant particles, as well as the quantification of experimentally accessible entanglement~\cite{Wiseman2003,Wiseman2003fb,Dowling2006}.

A significant barrier exists to fulfilling the goal of using particle entanglement as a resource in condensed matter physics -- its behavior is little known, outside of a few examples of non-interacting or few-body systems \cite{Haque2009,Eckert2002,Shi2003,Zozulya2008}. 
In this paper, we present a method to calculate particle entanglement in interacting many-body systems of itinerant bosons in the continuum described by the general Hamiltonian,
\begin{equation}
    H = \sum_{i=1}^N \left({ - \frac{\hbar^2}{2m_i} \nabla_i^2 + U_i}\right) + \sum_{i<j} V_{ij},
\label{eq:ham}
\end{equation}
where $m_i$ is the mass of a particle located at continuum spatial position $\vec{r}_i$, $U_i$ is an external potential (such as a harmonic trapping potential) and $V_{ij}$ is any two-body interaction, short or long ranged.  
The method is based on a path integral ground state Monte Carlo (PIGS) algorithm~\cite{Ceperley1995,Sarsa2000}, which we show gives access to the R\'enyi entanglement entropies through a ``replica'' trick.  The numerical algorithm is completely scalable in particle number and dimension of the system, unlike entanglement entropy estimators based on the direct calculation of the reduced density matrix. 

Below, we demonstrate that the R\'enyi entanglement entropies can be calculated accurately in PIGS simulations of
one dimensional ($1D$) interacting bosons.
We confirm 
the logarithmic scaling of the particle entanglement in the number of particles of the system, and demonstrate that the prefactor scales with interaction strength, before saturating to unity in the strongly interacting limit.
Additionally, we show how the R\'enyi entropy for a single particle is bounded by the condensate fraction. 

{\it Particle entanglement and R\'enyi entropies --} 
To define the particle entanglement in a system with $N$ particles, one first chooses a bipartition by identifying a $n$ particle subset.
Then, the entanglement can be quantified through the R\'enyi entropies,
\begin{equation}
S_{\alpha}(n) = \frac{1}{1-\alpha} \log \left({ {\rm Tr} \rho_n^{\alpha} }\right),
\end{equation}
where $\rho_n$ is the $n$-particle reduced density matrix on the ``$A$'' bipartition of $n$ particles,
with degrees of freedom associated with the $N-n$ particles in ``$B$'' traced out~\cite{Zanardi2002,Shi2003,Fang2003,Zozulya2008,Haque2009}.
Using this definition, we 
develop an algorithm for the calculation of particle entanglement in systems of itinerant bosons in the continuum, for integer R\'enyi entropies with $\alpha \ge 2$, 
based on the scalable simulation methodology of PIGS.
As a demonstration,  we calculate the single-particle entropy $S_2(n=1) = - \log {\rm Tr} \rho_1^2$ for interacting bosons on a ring.
Remarkably, in itinerant boson systems, the condensate fraction $n_0$ provides bounds on $S_2(n=1)$.
Much previous work has discussed the relationship between spatial entanglement and $n_0$~\cite{Simon2002,Vedral2003,Hines2003,Heaney2007,Heaney2007a,Kaszlikowski2007,Vedral2008,Heaney2009,Goold2009,Ding2009,Gagatsos2012,Calabrese2011a,Calabrese2011b,Burnett2001,Dunningham2002a}, but here we focus on particle entanglement.
First, $S_2$ is bounded from below by the single copy entanglement $S_{\infty}$, and above by $2S_{\infty}$.
Importantly,  
in itinerant boson systems,
this largest eigenvalue of $\rho_1$ is simply 
 $n_0$~\cite{Penrose:1956ve}, hence $S_{\infty} = - \log n_0$.
Next, we can place a tighter bound on $S_2$ by considering the two mode limit, where there are only two single particle 
modes accessible to the system. In this case, the restriction on ${\rm Tr} \rho$ means that the condensate fraction uniquely determines
$S_2$, since the binary entropy
$S_{\mathrm{bin}} = -\log (n_0^2 + (1-n_0)^2)$.  The final 
bounds for $n_0 > \frac{1}{2}$ are:
\begin{equation}
S_{\infty} \leq S_{\mathrm{bin}} \leq S_2 \leq 2 S_{\infty}, \label{bounds}
\end{equation}
with $S_{\infty}$ and $S_{\mathrm{bin}}$ switched for $n_0 \geq 1/2$.

{\it Computing entanglement entropy --} 
PIGS is a powerful and widely used method to study ground state 
properties of strongly interacting many-body systems~\cite{Ceperley1995,Sarsa2000,Cuervo2005a}. In the case of interacting bosons, its polynomial scaling 
allows for the study of large-scale systems in any dimension with short or long ranged interactions.  Since the recent demonstration that R\'{e}nyi entanglement entropy can be computed in Monte Carlo simulations~\cite{Hastings2010,Melko2010}, 
a large volume of subsequent effort has 
studied spatial entanglement entropy in lattice systems with quantum Monte Carlo (QMC).
To date however, the only QMC method that allows for the computation of entanglement entropy in a continuous-space system is a variational Monte Carlo method for fermions~\cite{McMinis2013}; no such methods have been reported for the calculation of particle entanglement. 

Following the approach of Ref.~\cite{Hastings2010}, we consider a ``replicated'' Hilbert space of a continuous-space system of $N$ bosons in first quantized notation. We represent a basis state as $\ket{\R}$, where $\R = \{\vec{r}_0,...,\vec{r}_{N-1}\}$ is a vector of all the particle positions in first quantized notation.
We label a 2nd copy of the Hilbert space $\{\ket{\tilde{\R}}\}$, and form a doubled, tensor product Hilbert space $\{\ket{\R;\tilde{\R}}\}$, these two replicas represent the same physical system and are non-interacting, giving us access to $S_2$.  For $S_{\alpha}$, one simply makes $\alpha$ non-interacting replicas.

To compute a bipartite R\'{e}nyi entropy, we must first define a subsystem by a particular choice of bipartitioning: for particle entanglement we choose a subset of $n$ particles $A$, such that $\R = \{\R_A,\R_B\}$. Due to the bosonic symmetry, any physical properties of this bipartition will not depend on the labels chosen for the subset, merely the number of particles $n$ in $A$. We define a permutation operator $\Pi_{\alpha}^A$ that maps $\R_A$ from one replica to another, modulo $\alpha$, and acts as the identity on all $\R_B$.  In the case of the second R\'enyi entropy \footnote{For $S_2$, $\Pi_{2}^A$ is called the ``SWAP'' operator in the literature for spatial entanglement.  However, we do not use this notation here to avoid confusion with a {\it swap} update used in a continuous space worm algorithm.}, 
$\Pi_{2}^A$ then simply 
interchanges the subset $A$ and $\tilde{A}$ between the two subsystems:
\begin{equation*}
    \Pi_{2}^A \eket{  \{\R_A,\R_B\}; \{\tilde{\R}_{\tilde{A}},\tilde{\R}_{\tilde{B}}\} } = 
\eket{  \{\tilde{\R}_{\tilde{A}},\R_B\};  \{\R_A,\tilde{\R}_{\tilde{B}}\} }.
\end{equation*}
The expectation value of this permutation operator of state $\ket{\Psi}$ in the doubled Hilbert space is related to the 2nd R\'{e}nyi entropy of $\ket{\Psi}$, $S_2$~\cite{Hastings2010, Cardy}:
\begin{equation}
\ebra{\Psi,\tilde{\Psi}} \Pi_{2}^A \eket{ \Psi,\tilde{\Psi} } = e^{-S_2}.
\end{equation}
To compute the R\'{e}nyi entropy of the ground state, we must be able to sample this replicated Hilbert space and define an estimator for the permutation operator.

PIGS methods give access to ground state properties of a many-body system through statistically sampling the imaginary time propagator $e^{-\beta H}$. Given a trial wavefunction $|\Psi_{\mathrm{T}}\rangle$, in the large imaginary time limit $\beta\rightarrow\infty$, $e^{-\beta H} \ket{\Psi_{\mathrm{T}}}$ converges to the ground state, as long as it is not orthogonal to it. 
We note that unlike some other zero temperature QMC methods, such as Green function and diffusion Monte Carlo, there is no inherent bias in estimators from the choice of the trial wavefunction, so long as $\beta$ is chosen to be sufficiently large~\cite{Boninsegni2012}.

\begin{figure}
\begin{center}
\includegraphics[width=\columnwidth]{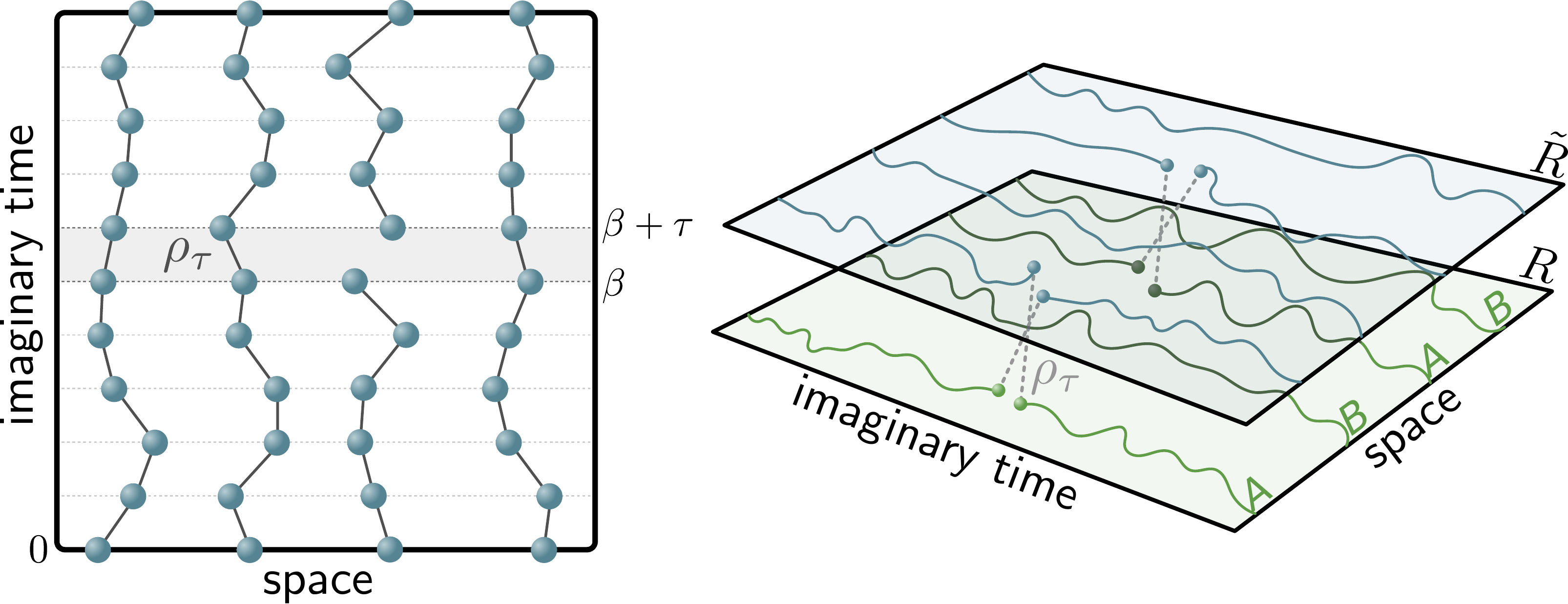}
\end{center}
\caption{(Color online). A configuration with $N=4$ bosons at zero temperature in
one spatial dimension with a single broken worldline (left). By creating $\alpha$ statistically equivalent replicas of the $N$ particle system ($\alpha=2$ here), each having $n$ broken worldlines, the $n$-particle R{\'e}nyi entropy can be measured via a permutation estimator (right).} 
\label{fig:worldlines}
 \end{figure}

To statistically sample observables, we consider a configuration space of imaginary time worldlines of the particles living in $d+1$ dimensions. In continuous space systems, we discretize the imaginary time direction and approximate the full propagator as a product of short time propagators $\rho_\tau = e^{-\tau H}$. The error in this approximation is controlled by the size of $\tau$, and
can be made smaller than any statistical uncertainties. We represent these discrete imaginary time worldlines by connected ``beads'' which represent a particle position at a given imaginary time as depicted in \figref{fig:worldlines}; links between these beads represent the short time propagator that relates a particle's position at neighboring imaginary times. Bosonic exchange symmetry is enforced by ensuring that both the trial wavefunction and any estimators
are symmetric over the fictitious particle labels.

To compute off-diagonal observables such as the $\Pi_2^A$ operator, we generate configurations that have $n$ open and $N-n$ closed worldlines, where the breaks are at the central imaginary time slice between $\beta$ and $\beta + \tau$ (\figref{fig:worldlines}). The broken worldlines of bipartition $A$ allow for the insertion of an off-diagonal operator with non-vanishing weight, while only closed worldlines belonging to bipartition $B$ are connected by a propagator $\rho_{\tau} ( \R^\beta_B; \R^{\beta+\tau}_B) = \bra{\R^\beta_B} \rho_\tau \ket{\R^{\beta+\tau}_B}$.
Such an ensemble may be Monte Carlo sampled by standard path integral methods using a variety of updates to ensure detailed balance~\cite{Ceperley1995}. 
The estimator for the $\Pi_2^A $ operator corresponds to sampling the statistical weight linking the worldlines of the $A$ particles with $\tilde{A}$ particles across the central time slice, as illustrated in \figref{fig:worldlines}. In this ensemble, the estimator for the permutation operator is,
\begin{equation}
\left \langle \Pi_2^A \right \rangle = 
\frac{\left \langle\rho^A_{\tau} \Biggl( \R^{\beta},\tilde{\R}^{\beta};\Pi_{2}^A \left(\R^{\beta+\tau},\tilde{\R}^{\beta+\tau}\right)\Biggr)\right \rangle }{\left \langle\rho^A_{\tau} \left( \R^{\beta},\tilde{\R}^{\beta};\R^{\beta+\tau},\tilde{\R}^{\beta+\tau}\right)\right \rangle }, 
\label{eq:SWAP}
\end{equation}
where we have defined the ``reduced propagator" $\rho_\tau^A \equiv \rho_\tau(\R^\beta;\R^{\beta +\tau})/ \rho_\tau(\R_B^\beta;\R_B^{\beta +\tau})$. The numerator in Eq.~\eqref{eq:SWAP} corresponds to the statistical weight of the ``permuted'' path and the denominator is a normalization factor arising from weight of the paths under the identity permutation. This form of the estimator is independent of the choice of the short-time propagator, which in general will involve diagonal weights at each bead as well as off-diagonal weights for the links \footnote{We have chosen to absorb the diagonal weights of the short time propagator in the estimator into the ensemble weights to improve efficiency \cite{algorithm}}. 

We note that the free particle contribution to the short time propagator in Eq.~\eqref{eq:SWAP} is a product of $n$ Gaussian factors for both $A$ and $\tilde{A}$. Consequently both the numerator and denominator in Eq.~\eqref{eq:SWAP} will be exponentially small in the size of the bipartition $n$; this is analogous to the exponential decay of the ``SWAP'' operator for spatial entanglement due to the ``area law'' in lattice systems~\cite{Hastings2010}. 
To address this issue for large $n$, we can perform a generalized ratio sampling, 
by building up the $n$-particle entanglement from a series of calculations of the $1-,2-,\ldots,(n-1)-$particle entanglement~\cite{Hastings2010}.
To compute spatial entanglement entropies, one may use a related method, where the broken worldlines only occur in a spatially defined region~\cite{algorithm}.

{\it Interacting bosons on a ring --} 
We present calculations of the 2nd R\'enyi single-particle entropy for the ground states of 
a system of $N$ interacting bosons in one spatial dimension with periodic spatial boundary conditions. The particles have a
repulsive 2-body interactions described by: $ V(r) = (2c/\sqrt{2 \pi \sigma^2})\exp[-r^2/2\sigma^2]$, which has integrated strength $2c$ in the thermodynamic limit and we have fixed the variance to be unity. As $\sigma \to 0$ we recover the Lieb-Liniger model of delta-function interacting bosons \cite{Lieb:1963ik}.  
Our simulations were performed at a constant density of $1/(5\sigma)$ for $N=2-16$, with $\hbar^2/(2 m \sigma^2)\simeq 6.06~\mathrm{K}$, ($k_\mathrm{B}=1$) and with an imaginary time path length $\beta$ chosen to ensure convergence within statistical errors using an $\mathcal{O}(\tau^4)$ approximation to the short time propagator \cite{Jang2001} with $\tau = \SI{0.08}{K^{-1}}$\footnote{The energy gap decreases with $N$ and increases with $c$, so the required $\beta$ varied from $2-12~K^{-1}$}.

\begin{figure}[t]
\begin{center}
\includegraphics[width=\columnwidth]{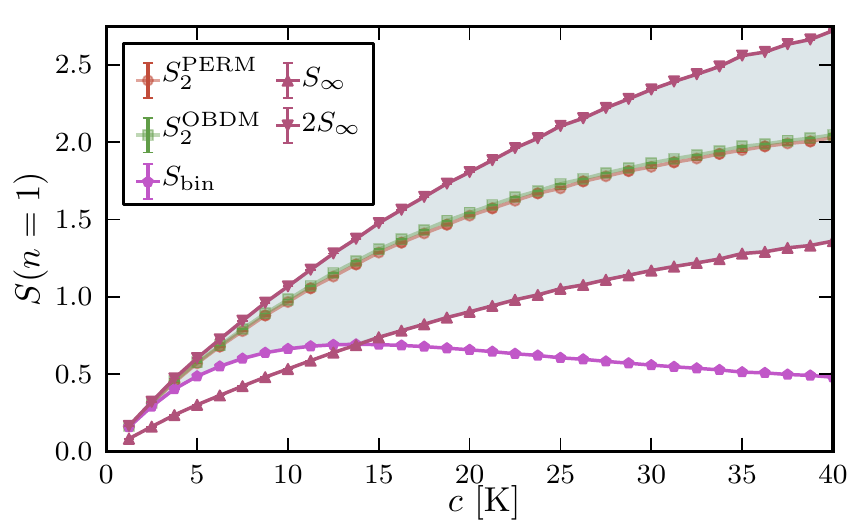}
\end{center}
\caption{(Color online). The single-particle entanglement entropy vs. interaction strength $c$, for $N=8$ bosons in $1D$. The 2nd R\'{e}nyi entropy $S_2$ computed via the permutation operator ($S_2^{\mathrm{PERM}}$) agrees exactly with that calculated from an integration of the one-body density matrix ($S_2^{\mathrm{OBDM}}$). The shaded region is the bound placed on $S_2$ from the condensate fraction alone. The lower bounds $S_{\mathrm{bin}}$ and $S_{\infty}$ cross when $n_0=1/2$.
} \label{fig:Svsc}
\end{figure}

The evaluation of the single-particle $S_2$ via the replica-trick
in this system allows for an important benchmark, as $S_2$ can also be calculated through numerical integration of the one-body density matrix $\rho_1$, which is accessible in PIGS. 
However, this latter procedure is {\it not} scalable as a function of spatial dimension or particle bipartition $n$, in contrast to the replica-trick calculation.  \figref{fig:Svsc} illustrates
the agreement of these two methods for computing $S_2$ in a $N=8$ system,
providing a crucial validation of our algorithm given in the previous section \footnote{Computing $S_2$ from numerical integration of $\rho_1$ also suffers from a systematic error due to discrete integration; here this systematic error has been reduced to be of order of the statistical errors.}.

Next, the zero-momentum component of $\rho_1$ is used to calculate the condensate fraction $n_0$, (which is non-zero in $1d$ only due to a finite size simulation cell), giving us access to $S_{\infty}$ needed for evaluation of the bounds in Eq.~(\ref{bounds}).  In Fig.~\ref{fig:Svsc} we see that $S_2$ is indeed bounded as expected.  Namely, in the weak interaction regime with a small condensate depletion ($n_0>1/2$),
$S_{\rm bin}$ and $2S_{\infty}$ provide tight bounds for $S_2$.  In contrast, for $n_0 \le 1/2$,  $S_{\rm bin}$ saturates at a maximum of $\log 2$ and then decreases, whereas $S_{\infty}$ diverges logarithmically as $n_0 \rightarrow 0$, providing a tighter lower bound for $S_2$ in this regime.

Finally, we examine the scaling of the single-particle entanglement.  At fixed interaction strength and density, there are two scaling parameters, $n$ and $N$.
Whereas the ``area law'' of spatial entanglement arises from scaling the entropy with bipartition
size, for particle entanglement the bipartition is solely characterized by the number of particles in the subregion, $n$. A canonical form of this scaling was proposed by Zozulya, Haque, and Schoutens  \cite{Zozulya2008} to be $S(n,N) = a n\log N + b$, and
has been derived for various limiting cases \cite{Haque2009}.

\begin{figure}[t]
\begin{center}
\includegraphics[width=\columnwidth]{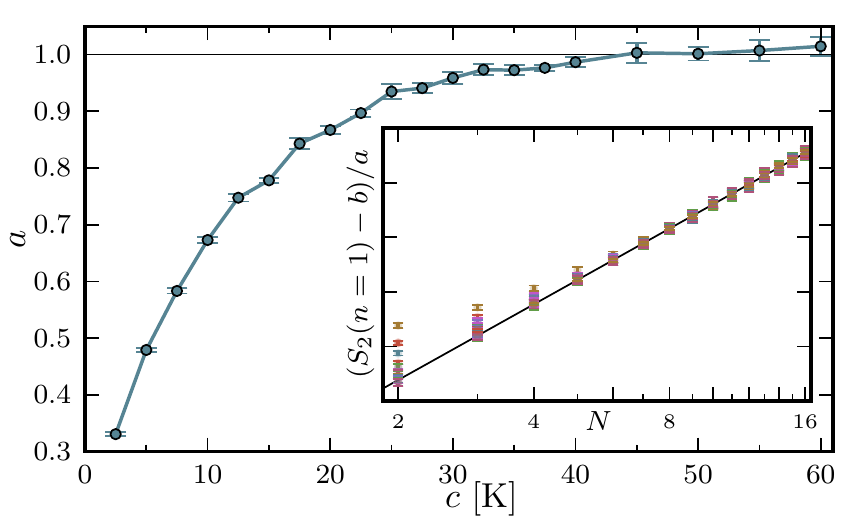}
\end{center}
\caption{(Color online). The coefficient of the $\log N$ term in the finite-size scaling of $S_2(n=1)$, $a$, vs. interaction strength $c$ for $N=2-16$. The saturation to unity occurs where the interactions become effectively hard-core; $a<1$ indicates multiple occupancy of single-particle modes. {\it Inset:} Data collapse of the finite-size scaling of $S_2(n=1)$ with the solid line equal to $\log N$. Different colors correspond to the different interaction strengths plotted in the main figure.  }
   \label{fig:avsc}
 \end{figure}

The inset of \figref{fig:avsc} illustrates the data collapse of $S_2(n=1)$ to the $\log N$ scaling form for $N=2-16$ and $c=2.5-\SI{60}{K}$, confirming 
it in a system with non-trivial interactions. The main plot in \figref{fig:avsc} shows the scaling of the coefficient $a$, which saturates to unity in the strongly interacting limit. This scaling of $a$ is qualitatively consistent with that of the single particle entanglement of a Luttinger liquid (by which this system should be described in the thermodynamic limit) where $a$ is the inverse of the Luttinger parameter.
Note, the coefficient of the scaling with the bipartition size is given by $a \log N$.  Unlike spatial entanglement where the coefficient of the area law generally arises due to short distance physics, the analogous coefficient for particle entanglement is logarithmically extensive. A logarithmic scaling also appears in the spatial entanglement of itinerant particles, where its origin is number fluctuations in the spatial bipartition~\cite{Simon2002,Vedral2003,Ding2009,Calabrese2011a,Calabrese2011b,Song2010}. The appearance of $\log N$ in the particle entanglement is distinct from the spatial entanglement case and  arises when there is no multiple occupancy of single particle modes (e.g.~in hard-core bosons or fermions). This is a consequence of the fact that the single particle entanglement measures the effective number of accessible single particle modes.

{\it Discussion --} 
We have presented a scalable simulation method for computing particle entanglement entropies in continuum systems of itinerant bosons, 
and implemented it in ground state path integral Monte Carlo.
This technique is applicable to a wide range of models which can directly address the physics of experimentally relevant quantum fluids and gases.
Through simulations of interacting bosons on a ring, 
we have demonstrated how the condensate fraction bounds the particle entanglement, allowing for the possibility of quantifying entanglement experimentally. We have also verified a logarithmic scaling of particle entanglement with particle number for a system with non-trivial interactions. 

Future studies should include a study of the bipartition size scaling, providing 
access to the particle entanglement analog of the ``area law''. Algorithmic extensions of our method may also be used to compute the spatial entanglement entropy in continuum systems.  The ability to address experimentally relevant models opens up the possibility of quantifying {\it accessible} entanglement in quantum information systems~\cite{Wiseman2003,Wiseman2003fb,Dunningham2005,Dowling2006}, as well as superfluid helium-4, cold atomic gases, and doped parahydrogen clusters~\cite{Li2010}.  In addition to characterizing quantum phases of matter, knowledge of particle entanglement may yield a deep understanding of classical simulability of quantum systems of itinerant particles~\cite{Verstraete2010}, as has been the case for spatial entanglement in lattice systems~\cite{Latorre2007,Schuch2008a}.

{\it Acknowledgments --}
We thank E.~M.~Stoudenmire and S.~Inglis for enlightening discussions and
acknowledge the use of the computing facilities of Compute Canada (RQCHP's Mammouth cluster) and the Vermont Advanced Computing Core supported by NASA (NNX-08AO96G).
This research was supported by NSERC of Canada, the Perimeter Institute for Theoretical Physics, the John Templeton Foundation, and the University of Vermont.

\bibliographystyle{apsrev4-1.bst}
\bibliography{refs}

\end{document}